# Optical freezing of charge motion in an organic conductor

Takahiro Ishikawa[1], Yuto Sagae[1], Yota Naitoh[1], Yohei Kawakami[1],

Hirotake Itoh[1,2], Kaoru Yamamoto[3], Kyuya Yakushi[4], Hideo Kishida[5],

Takahiko Sasaki[6], Sumio Ishihara[1], Yasuhiro Tanaka[7], Kenji Yonemitsu[7],

and Shinichiro Iwai[1,2*]

[1]Department of Physics, Tohoku University, Sendai 980-8578, Japan

[2]JST, CREST, Sendai 980-8578, Japan

[3]Department of Applied Physics, Okayama Science University, Okayama, 700-0005 Japan

[4]Toyota Physical and Chemical Research Institute, Nagakute, 480-1192, Japan

[5]Department of Applied Physics, Nagoya University, Nagoya 464-8603, Japan

[6]Institute for Materials Research, Tohoku University, Sendai 980-8577, Japan

[7]Department of Physics, Chuo University, Tokyo 112-8551, Japan

71.27.+a, 75.30.Wx, 78.47.jg

*s-iwai@m.tohoku.ac.jp

Abstract

Dynamical localization, i.e., reduction of the intersite electronic transfer integral $t$ by an alternating electric field, $E(\omega)$, is a promising strategy for controlling strongly correlated systems with a competing energy balance between $t$ and the Coulomb repulsion energy. Here we describe a charge localization induced by the 9.3 MV/cm instantaneous electric field of a 1.5 cycle (7 fs) infrared pulse in an organic conductor α-(bis[ethylenedithio]-tetrathiafulvelene)$_2$I$_3$. A large reflectivity change of > 25% and a coherent charge oscillation along the time axis reflect the opening of the charge ordering gap in the metallic phase. This optical freezing of charges, which is the reverse of the photoinduced melting of electronic orders, is attributed to the ~10% reduction of $t$ driven by the strong, high-frequency ($\omega \geq t/\hbar$) electric field.

## Introduction

Ultrafast control of conduction and magnetic properties in strongly correlated systems[1-3] has been extensively studied from the perspective of photoinduced insulator-to-metal transitions, or, equivalently, the "melting" of electronic orders [red arrow in Fig. 1(a)] in Mott insulators,[4-10] charge ordered systems[11-14], and charge/spin density wave materials.[6, 15-20] Recently, the excitation of coherent phonons has been established as a leading strategy for the melting/constructing of electronic orders.[8, 9,15-20] On the other hand, the development of strong electric fields (> MV/cm) of few-cycle optical pulses and recent theoretical studies using dynamical mean-field theory (DMFT) suggest that extreme non-equilibrium electronic states such as Floquet states, negative temperatures, and superconducting states[21,22] can be achieved.

In highly non-equilibrium phenomena, reducing the intersite transfer integral $t$ by modulating the site energy under strong continuous-wave (CW)[21,23-26] and pulsed[22,26] alternating current (AC) electric fields $E(\omega)$, shown in Fig. 1(b), plays an important role. Such an intense modulation of the electronic structure driven by a strong electric field, referred to as "dynamical localization", provides a new strategy for controlling charge motion in strongly correlated materials, which have a competing energy balance between the on-site or intersite Coulomb repulsion and $t$ (ref. 27).

Our target material is a layered organic conductor, α-(ET)$_2$I$_3$ (ET: bis[ethylenedithio]-tetrathiafulvelene), which exhibits a thermal (equilibrium) metal-to-charge-ordered (CO) insulator transition at $T_{CO}$ = 135

K,[28-32] and a photoinduced (non-equilibrium) transition [14] as illustrated in Fig. 1(a). Efficient photoinduced insulator-to-metal transitions of > 200 ET molecules/photon have previously been demonstrated.[14] However, the optical response of the metallic phase at $T > T_{CO}$ remains unclear.

In this study, we perform pump-probe transient reflectivity measurements for the metallic phase of α-(ET)$_2$I$_3$ using 7 fs 1.5 cycle infrared pulses. Our results demonstrate the optical freezing of charges or, equivalently, photoinduced charge localization in the metallic phase [blue arrow in Fig. 1(a)]. We discuss the mechanism on the basis of theories that have been proposed for non-equilibrium states generated by high-frequency CW and pulsed AC fields.

## Results

### Thermal metal-to-charge ordered insulator transitions.

The optical conductivity in the mid- and near-infrared region in Fig. 2(a) has a clear opening of the CO gap at ~0.1–0.2 eV and a spectral weight transfer to a higher energy at the thermal metal-to-insulator transition.[29,30] Figure 2(b) shows the steady state reflectivity ($R$) at 140 K (metal: solid red curve) and 40 K (CO: dashed blue curve). As shown in Fig. 2(d), at $T_{CO}$ $R$ abruptly decreases for 0.09 eV (open circles) and increases for 0.64 eV (closed circles) as the temperature decreases. Such a large change (80% at 0.64 eV, 60% at 0.09 eV) in $R$ at the metal-to-insulator transition directly corresponds to the opening of the CO gap and the transfer of the spectral weight to a higher energy of the optical conductivity,[29,30] as shown in Fig. 2(a). Figure 2(c) shows the spectral differences at different temperatures. The solid curve shows [$R$(40 K) - $R$(140 K)]/$R$(140 K), reflecting the metal-to-insulator change across $T_{CO}$. The dashed, and dashed-dotted curves show [$R$(190K) - $R$(140 K)]/$R$(140 K), and [$R$(170K) - $R$(140 K)]/$R$(140 K), exhibiting increases in the electron-lattice temperature up to 190 K and 170 K, respectively. Thus, the marked increase of $R$ at > 0.55 eV (blue shading) clearly characterizes the metal-to-insulator change across $T_{CO}$, while the rise of the electron-lattice temperature is detected as a reflectivity decrease below 0.65 eV.

**Photoinduced charge localization.** Figure 3(a) shows the transient reflectivity ($\Delta R/R$) spectra for time delays ($t_d$) of 30 fs (closed blue circles for an excitation intensity $I_{ex}$ = 0.8 mJ/cm$^2$, open blue circles for $I_{ex}$ = 0.12 mJ/cm$^2$) and 300 fs (closed black circles for $I_{ex}$ = 0.8 mJ/cm$^2$) after excitation by a 7 fs pulse covering the spectral range 0.6–0.95 eV. The spectrum shown by crosses represents $\Delta R/R$ at $t_d$ = 300 fs after a 100-fs pulse excitation for $I_{ex}$ = 0.8 mJ/cm$^2$. A large increase of $R$ ($\Delta R/R$ ~ 0.28 at 0.66 eV) was observed at $t_d$ = 30 fs for $I_{ex}$ = 0.8 mJ/cm$^2$, as shown by the blue shading in Fig. 3(a). The spectral shape of $\Delta R/R$ is analogous to that of the temperature differential spectrum [$R$(40K) - $R$(140 K)]/$R$(140 K) in Fig 2(c), suggesting that a photoinduced metal-to-insulator change occurred. In contrast, as shown by the crosses, the black closed circles, and the red shading, the spectral shapes and the magnitude of $\Delta R/R$ measured at $t_d$ = 300 fs are analogous to the differential spectrum [$R$(190K) - $R$(140K)]/$R$(140K), reflecting an increase in the electron-lattice temperatures up to 190 K. In such a time domain (300 fs), the lattice system cannot be thermalized and a quasi-thermalized state is formed as a result of the interaction between the charge and various short-period vibrational and optical phonon modes. Therefore, the lattice temperature shown in the reflectivity spectrum at $t_d$=300 fs can be defined as a local temperature consisting solely of short-period modes. For $I_{ex}$ = 0.12 mJ/cm$^2$, the $\Delta R/R$ spectrum shown by the open blue circles indicates that the metal-to-charge-localized change does not occur under weak excitation conditions.

**Charge ordering gap oscillation along the time axis.** The temporal change of the $\Delta R/R$ spectrum is plotted as a 2-dimensional (probe energy– delay time) map in Fig. 3(b). Positive and negative $\Delta R/R$ are shown by the blue and red shadings, respectively. The spectrum for $t_d$ < 50 fs, reflecting the transient CO state, oscillates with a period of 20 fs as indicated by the red dotted lines. Then, the spectral shape of $\Delta R/R$ markedly changes with the melting of the transient CO and the increase of electron/lattice temperatures shown as the red area in the time scale of 50–100 fs. Such spectral change occurs simultaneously as the oscillation decays. The time profile of $\Delta R/R$ sliced at 0.64 eV is shown in Fig. 4(a). A positive $\Delta R/R$ (solid curve with blue shading) at $I_{ex}$ = 0.8 mJ/cm$^2$ persists for $t_d$ ~ 50 fs after the excitation pulse. Then, $\Delta R/R$ becomes negative (red shading), indicating that the photoinduced CO state collapsed because of the increase of electron-lattice temperature. In contrast, for $I_{ex}$ = 0.12 mJ/cm$^2$, a positive signal was not detected at 0.64 eV, as shown by the dashed-dotted curve. It is worth noting that the time profile was modulated by the oscillating component with a period of 20 fs. This oscillation, shown in Fig. 4(b), can be attributed to the intermolecular charge oscillation reflecting the CO gap, because the time-resolved spectrum of the oscillating component at $t_d$=0–40 fs obtained by the wavelet (WL) analysis [blue curve, inset of Fig. 4(b)] corresponds to the optical conductivity spectrum near the CO gap of ~0.1 eV at 10 K(black curve in the inset).

Although the C=C vibration energy is near the CO gap,[29,30] the vibrational contribution can be detected separately after $t_d$ >50 fs, i.e., the WL spectra at $t_d$ =80–120 fs indicated by the red curve in the inset of Fig. 4(b) shows a dip at ~0.15 eV (which equals the vibration peak of the optical conductivity at 140 K[29,30] in the inset) reflecting the destructive interference between charge motion and vibration.[33] Therefore, the intense 20 fs oscillation detected before the appearance of the charge-vibration interference dip is attributed to the oscillation of the CO gap. Such CO gap oscillation has been detected in the precursor step to the CO melting,[33] although the amplitude was much smaller (<1 % of the $\Delta R/R$ signal) than the present case. Because of the ultrafast decay within 50 fs (corresponding to the energy scale of $\hbar/50 fs$ =0.08 eV) of the photoinduced state, we cannot verify the insulating gap below 0.1 eV. However, the gap-like spectral shape at 0.1–0.2 eV, as described above, clearly indicates that the charge motion on the corresponding energy scale is frozen as if the charge distribution on the ET molecules were analogous to that in the CO insulator. In this limited sense, the CO gap is regarded as opened.

Considering that the charge-vibration interference dip appears after the photoinduced charge localization occurs, the vibration-induced mechanism is ruled out, i.e., the charge localization is driven by the electronic interaction, although vibration may play some role in stabilizing the transient CO state. Here, the lifetime of the transient CO state (<50 fs) is much shorter than that of the coherent phonon- induced state (~ 1 ps or longer) which has been reported as the melting of a Mott insulator[8,9] or the ordering of the transient spin density wave.[19, 20] Such a short lifetime indicates that the transient CO state is not adequately stabilized by vibrational and lattice motions.

From $\Delta R/R = 0.28$ at 0.64 eV and the temperature difference $[R(40K) - R(140 K)]/R(140 K) = 0.8$ at the same energy, the volume fraction of the photoinduced CO state is evaluated to be ~35%. As $I_{ex} = 0.8$ mJ/cm$^2$ corresponds to ~0.0145 photons/ET molecule, the efficiency of this process is 25 ET/photon. This is much lower than that for photoinduced melting of the CO state(>200 ET/photon[14]), demonstrating that the mechanism of the photoinduced charge localization is completely different from that of the insulator-to-metal transition.

**Discussion.**

Figure 5(a) shows $\Delta R/R$ at 0.64 eV as a function of the instantaneous electric field $E_0$ ($\propto \sqrt{I_{ex}}$) for various temperatures. An increase in $R$ (positive $\Delta R/R$) is detectable only for $I_{ex} > 0.31$ mJ/cm$^2$ ($E_0 = 5.8$ MV/cm) at 138 K, whereas the negative $\Delta R/R$ for $I_{ex} < 0.31$ mJ/cm$^2$ is due to the rise in the electron-lattice temperature. As shown in the inset, the efficiency of the photoinduced charge localization becomes larger near $T_{CO}$. We can evaluate the positive component of $\Delta R/R$ by subtracting the negative component of the time profile shown in Fig. 4(a), assuming that the negative component grows exponentially along the time axis [the dashed red curve in Fig. 4(a)]. The resultant positive component exhibits a non-linear increase as a function of $E_0$, as shown in Fig. 5(b).

In non-equilibrium states induced by a strong AC field $\mathbf{E}(\omega)$, $t$ is reduced for CW[21,23-25] and pulsed[22,26] light. According to the dynamical localization theory[23-26] for a CW AC field, the effective $t$ (= $t'$) can be represented as,

$$t' = t_0 J_0 (\Omega_{AB}/\omega), \qquad (1)$$

where $t_0$ is the transfer integral for zero field. Here, $\Omega_{AB} \equiv e\mathbf{r}_{A,B} \cdot \mathbf{E}_0/\hbar$, where $\mathbf{E_0}$ and $\omega$, respectively, denote the amplitude and angular frequency of the electric field of the light; $\mathbf{E}(t) = \mathbf{E}_0 \sin(\omega t)$; $\mathbf{r}_{A,B}$ is the vector from the A site to its nearest-neighbour B site of α-(ET)$_2$I$_3$, as shown in Fig. 1(a) ; and $e$ is the elementary charge. Equation is satisfied for CW light. In addition, on the basis of the numerical solution of the time-dependent Schrödinger equation for the two-site tight-binding model and a model for a CO molecular conductor,[26] this relation has recently been shown to be satisfied by pulsed light, with respect to the efficiency of the intersite electronic transition (i.e., the energy increment due to the pulsed light). Also, in the DMFT calculations, the change in the electronic state into one with modified $t$ appears within the time scale of a few optical cycles after the sudden application of a CW AC field.[21] This fact also suggests that the dynamical localization functions transiently with the few-cycle pulse. If we use the parameter $r_{AB}\cos\theta$ =5.4 angstroms (refs 34,35); where $\theta$ represents the relative angle between $\mathbf{r}_{A,B}$ and $\mathbf{E}$, $t$ is reduced, as it is proportional to the zeroth order Bessel function $J_0$, and becomes zero at 40 MV/cm ($\Omega_{AB}/\omega = 2.40$), as shown in Fig. 5(c). From Fig. 5(c), we can estimate that the change in $J_0 \propto t'$ is approximately 10% for a typical instantaneous field of 9.3 MV/cm ($I_{ex}$ = 0.8 mJ/cm²).

To demonstrate the instability of the metallic phase induced by this 10% change in $t$, we roughly estimated the change of $T_{CO}$ by changing $t$. Figure 6(a) shows the temperature dependence of the hole densities ($\rho_H$) at the A and A' molecules in Fig. 1(a) as a function of the normalized temperature $T/T_{CO}$, calculated using the Hartree–Fock approximation for an extended Hubbard model.[36] A charge disproportionation occurs below $T/T_{CO}$ = 1, as indicated by the closed (A: charge rich) and open (A': charge poor) circles for the original $t_0$. We also calculated the $\rho_H$ vs. $T/T_{CO}$ relation with changing $t$, where the rectangles, upward triangles, and downward triangles show the calculated values for $0.95t_0$, $0.9t_0$, and $0.85t_0$, respectively. Figure 6(b) illustrates that the change in $T_{CO}$ (= $\Delta T_{CO}/T_{CO}$) is proportional to the decrease in $t$ (=$-\Delta t/t$), indicating that a 10% change in $t$ causes a 12% change in $T_{CO}$ from 135 K to 152 K. Thus, $T_{CO}$ increases across the measured temperature (138 K).

For our experimental conditions (< 12 MV/cm), $1/t'$ (in units of $1/t_0$) shows a non-linear dependence on $E_0$, [red curve in Fig. 5(b)]. This non-linear dependence of $1/t'$ on $E_0$ is consistent with the relation between $\Delta R/R$ and $E_0$ [closed circles in Fig. 5(b)]. Considering that $R$ reflects the metal-to-insulator transition [Fig. 2(b)], it is reasonable to assume that $\Delta R/R$ is proportional to the efficiency of the transient charge localization. As such, the $E_0$ dependence of $\Delta R/R$, can be described by $1/t'$.

According to recent studies on sub-cycle asymmetric pulses,[22] the momentum shift of the band structure is given by the "dynamical phase" $\phi = \int dt E(t)$ (ref. 22). However, in the present case, this is estimated to be $\phi=1.66\times10^{-4}\,[rad]$ at most, which is much smaller than $\phi=\pi/2\,[rad]$, and is too small to cause a detectable momentum shift. Such shift would be detected if the width of the asymmetric pulse were shorter than ~3 fs at this wavelength. Another problem lies in the mechanism driving the reduced $t$, which should be applicable to events before the excitation ends, i.e., ~7 fs. The reason why the photoinduced CO persists for ~50 fs after the 7 fs pulse is unclear at the moment. Further experiments and theoretical considerations will be needed to clarify this issue.

In summary, this report demonstrates a large reflectivity increase (> 25%) and a coherent CO gap oscillation along the time axis indicating the opening of a CO gap by the 9.3 MV/cm electric field of a 1.5 cycle, 7 fs near-infrared pulse in an organic conductor α-(ET)$_2$I$_3$. The plausible mechanism for such a dramatic change in the electronic state is the reduction of $t$ (~10%) driven by this strong high-frequency field.

# Methods

**Sample preparation.** Single crystals of $\alpha$-(ET)$_2$I$_3$ (2 × 1 × 0.1 mm) were prepared using a method described in a previous study.[28]

**7 fs infrared pulse generation.** A broadband infrared spectrum for the 7 fs pulse covering 1.2–2.3 μm, shown by the orange curve in Fig. 2(a), was obtained by focusing a carrier-envelope phase (CEP) stabilized idler pulse (1.7 μm) from an optical parametric amplifier (Quantronix HE-TOPAS pumped by Spectra-Physics Spitfire-Ace) onto a hollow fibre set within a Kr-filled chamber (Femtolasers). Pulse compression was performed using both active mirror (OKO technologies 19-ch linear MMDM) and chirped mirror (Femtolasers, Sigma-Koki) techniques. The pulse width derived from the autocorrelation of the generated second harmonic was 7 fs, which corresponds to 1.5 optical cycles. The instantaneous electric field on the sample surface (excitation diameter 200 μm) for a typical excitation intensity can be evaluated as

$$E_{peak} = \sqrt{\frac{2}{\varepsilon_0 c}}\sqrt{I_{peak}} = 27.4 \times \sqrt{1.14 \times 10^{11} (\text{W/cm}^2)} = 9.25 \times 10^6 (\text{V/cm})\ ,\ \text{where}\ I_{peak} =$$

1.14 × 10$^{11}$ W/cm$^2$ represents the peak power for an excitation intensity $I_{ex}$ of 0.8 mJ/cm$^2$.

**Transient reflectivity measurements.** We performed transient reflectivity experiments using both 7 fs and 100 fs pulses. The excitation photon energies for 7 fs and 100 fs pulses were 0.6–0.95 eV (7 fs) and 0.89 eV (100 fs), respectively. In the transient reflectivity measurement using a 7 fs pulse, the probe pulse reflected from the sample was detected by InGaAs detector (New-Focus model 2034) after passing through a spectrometer (Bunkoukeiki, M10). The pump-on and pump-off were alternately switched by the feed-back-controlled optical chopper (New Focus Model 3501), synchronized with the laser driver. Each probe shot was sampled using boxcar integrators (Stanford Research SR250). After normalization by a reference pulse, the observed intensity of respective shots were recorded in the PC.

Acknowledgements

We are grateful to T. Oka (University of Tokyo), and Y. Kayanuma (Tokyo Institute of Technology) for their insightful discussions. This work was supported by Grant-in-Aid for Scientific Research (A) No. 23244062.

Author contributions

T. I. and Y. K. developed the 7 fs light source and carried out the transient reflectivity measurements using this pulse. Y. S., Y. N. and H. I performed 100 fs transient reflectivity experiments. T. I., Y. S. and Y. N. analyzed the data. K. Yamamoto, K. Yakushi, H. K, and T. S. performed the synthesis and the characterization of the single crystal. S. Ishihara, Y. T. and K. Yonemitsu made theoretical considerations and calculations. S. Iwai devised all the experiments. T. I., S. Iwai, and K. Yonemitsu wrote the paper after discussing with all the co-authors.

**References**

[1] Koshihara, S. & Kuwata-Gonokami, M. (eds.) *Special Topics: Photo-Induced Phase Transitions and Their Dynamics*, *J. Phys. Soc. Jpn.* **75,** 011001-011008 (2006).

[2] Basov, D. N., Averitt, R. D., van der Marel, D., Dressel, M., & Haule, K. Electrodynamics of correlated electron materials. *Rev. Mod. Phys.* **83,** 471-541 (2011).

[3] Yonemitsu, K. & Nasu, K. Theory of photoinduced phase transitions in itinerant electron systems. *Physics Reports* **465,** 1-60 (2008).

[4] Cavalleri, A., et al. Femtosecond structural dynamics in $VO_2$ during an ultrafast solid-solid phase transition. *Phys. Rev. Lett.* **87,** 237401 (2001).

[5] Iwai, S., et al. Ultrafast optical switching to a metallic state by photoinduced Mott transition in a halogen-bridged nickel-chain compound. *Phys. Rev. Lett.* **91,** 057401 (2003).

[6] Perfetti, L., et al. Time evolution of the electronic structure of $1T$-$TaS_2$ through the insulator-metal transition. *Phys. Rev. Lett.* **97,** 067402 (2006).

[7] Okamoto, H., et al. Photoinduced metallic state mediated by spin-charge separation in a one-dimensional Organic Mott insulator. *Phys. Rev. Lett*, **98,** 037401(2007).

[8] Kawakami, Y., et al. Optical modulation of effective on-site Coulomb energy for the Mott transition in an organic dimer insulator. *Phys. Rev. Lett.* 066403(2009).

[9] Kaiser, S., et al., Optical properties of a vibrationally modulated solid state Mott insulator. *Sci. Rep.* 4, 3823(2014).

[10] Liu, M., et al. Terahertz-field-induced insulator-to-metal transition in vanadium dioxide metamaterial. *Nature* **487,** 345-348 (2012).

[11] Miyano, K., Tanaka, T., Tomioka, Y., & Tokura, Y. Photoinduced insulator-to-metal transition in a Perovskite manganite. *Phys. Rev. Lett.* **78,** 4257-4260 (1997).

[12] Chollet, M., et al. Gigantic photoresponse in ¼-filled-band organic salt (EDO-TTF)$_2$PF$_6$. *Science* **307,** 86-89 (2005).

[13] Polli, D., et al. Coherent orbital waves in the photo-induced insulator-metal dynamics of a magnetoresistive manganite. *Nature Mater.* **6,** 643-647 (2007).

[14] Iwai, S., et al. Photoinduced melting of a stripe-type charge-order and metallic domain formation in a layered BEDT-TTF-based organic salt. *Phys. Rev. Lett.* **98,** 097402 (2007).

[15] Demsar, J., Forró, L., Berger, H., & Mihailovic, D. Femtosecond snapshots of gap-forming charge-density-wave correlations in quasi-two-dimensional dichalcogenides 1*T*-TaS$_2$ and 2*H*-TaSe$_2$. *Phys. Rev.* B **66,** 041101(R) (2002).

[16] Hellmann, S., et al. Time-domain classification of charge-density-wave insulators. *Nature Commun.* **3,** 1069 (2012).

[17] Schmitt, F., et al. Transient electronic structure and melting of a charge density wave in TbTe$_3$. *Science* **321,** 1649-1652 (2008).

[18] Stojchevska, L., et al. Ultrafast switching to a stable hidden quantum state in an electronic crystal. *Science* **344,** 177-180 (2014).

[19] Kim, K. W., et al. Ultrafast transient generation of spin-density-wave order in the normal state of BaFe$_2$As$_2$ driven by coherent lattice vibrations. *Nature Mater.* **11,** 497-501(2012).

[20] Yang, L. X., et al., Ultrafast modulation of the chemical potential in BaFe$_2$As$_2$ by coherent phonons. *Phys. Rev. Lett.*, **112**, 207001(2014).

[21] Tsuji, N., Oka, T., Werner, P., & Aoki, H. Dynamical band flipping in fermionic lattice systems: An ac-field-driven change of the interaction from repulsive to attractive. *Phys. Rev. Lett.* **106,** 236401 (2011).

[22] Tsuji, N., Oka, T., Aoki, H., & Werner, P. Repulsion-to-attraction transition in correlated electron systems triggered by a monocycle pulse. *Phys. Rev. B* **85,** 155124 (2012).

[23] Dunlap, D. H. & Kenkre, V. M. Dynamic localization of a charged particle moving under the influence of an electric field. *Phys. Rev. B* **34,** 3625-3633 (1986).

[24] Grossmann, F., Dittrich, T., Jung, P., & Hänggi, P. Coherent destruction of tunneling. *Phys. Rev. Lett.* **67**, 516 (1991)

[25] Kayanuma, Y. & Saito, K. Coherent destruction of tunneling, dynamic localization, and the Landau-Zener formula. *Phys. Rev. A* **77**, 010101(R) (2008).

[26] Nishioka, K. & Yonemitsu, K. Intra- and interdimer transfer integrals effectively modified by pulsed and continuous-wave lasers for controlling charge transfers in molecular crystals. *J. Phys. Soc. Jpn.* **83,** 024706 (2014).

[27] Imada, M., Fujimori, A., & Tokura, Y. Metal-insulator transitions. *Rev. Mod. Phys.* **70,** 1039-1263 (1998).

[28] Bender, K., et al. Synthesis, structure and physical properties of a two-dimensional organic metal, di[bis(ethylenedithiolo)tetrathiofulvalene] triiodide, (BEDT-TTF)$^+$$_2$ I$^-$ $_3$. *Mol. Cryst. Liq. Cryst.* **108,** 359-371 (1984).

[29] Dressel, M., Grüner, G., Pouget, J. P., Breining, A., & Schweitzer, D. Field- and frequency dependent transport in the two-dimensional organic conductor α-(BEDT-TTF)$_2$I$_3$. *J. Phys. I France* **4,** 579-594 (1994).

[30] Yue, Y., et al. Nonuniform site-charge distribution and fluctuations of charge order in the metallic state of α-(BEDT-TTF)$_2$I$_3$. *Phys. Rev. B* **82,** 075134 (2010).

[31] Yamamoto, K., et al. Strong optical nonlinearity and its ultrafast response associated with electron ferroelectricity in an organic conductor. *J. Phys. Soc. Jpn.* **77,** 074709 (2008).

[32] Ivek, T., et al. Collective excitations in the charge-ordered phase of α-(BEDT-TTF)$_2$I$_3$. *Phys. Rev. Lett.* **104,** 206406 (2010).

[33] Kawakami, Y., et al. Early-stage dynamics of light-matter interaction leading to the insulator-to-metal transition in a charge ordered organic crystal. *Phys. Rev. Lett.* **105,** 246402 (2010).

[34] Mori, T., Mori, H. &Tanaka, S. Structural genealogy of BEDT-TTF-based organic conductors II. Inclined molecules: θ, α, and κ phases. *Bull. Chem. Soc. Jpn.* **72**, 179-197(1999).

[35] Kakiuchi, T. et al. Charge ordering in α-(BEDT-TTF)$_2$I$_3$ by synchrotron X-ray diffraction., J. Phys. Soc. Jpn., 76, 113702(2007).

[36] Tanaka, Y. & Yonemitsu, K. Charge order with structural distortion in organic conductors: Comparison between θ-(ET)$_2$RbZn(SCN)$_4$ and α-(ET)$_2$I$_3$. *J. Phys. Soc. Jpn.* **77,** 034708 (2008).

Figure Legends

Fig. 1 **Melting and freezing of charge motion in α-(ET)$_2$I$_3$ and AC-field-induced charge localization.**
**(a)** CO insulator-to-metal transition (red arrow, melting) and metal-to-CO insulator transition (blue arrow, freezing) for both thermal(equilibrium) and optical(non-equilibrium) transitions in α-(ET)$_2$I$_3$. **(b)** CO gap is opened by the decrease of transfer integral $t$, because of the competing energy balance between Coulomb interaction $V$ and $t$. $t$ is changed from $t_0$ to $t'$, induced by the strong AC electric field $E(\omega)$.

Fig. 2 **Steady state optical conductivity and reflectivity spectra.**
**(a)** Optical conductivity ($\sigma$) spectra of α-(ET)$_2$I$_3$ at 40 K (CO: dashed blue curve) and at 140 K (metal: solid red curve). The spectrum of the 7 fs pulse is indicated by the orange curve. **(b)** Reflectivity ($R$) spectra at 40 K (CO) and at 140 K (metal). **(c)** The spectra for three temperature differentials: [$R$(40K) - $R$(140K)]/$R$(140K) (solid curve with blue shading), [$R$(190K) - $R$(140K)]/$R$(140 K) [dashed curve (x3) with red shading], and [$R$(170K) - $R$(140K)]/$R$(140 K) [dashed-dotted curve (x3)]. **(d)** Reflectivities measured at 0.09 eV and 0.64 eV [indicated by the blue arrows on the spectrum in Fig. 2(b)], plotted as a function of temperature.

Fig. 3 **Transient reflectivity spectra and their time evolutions.**
**(a)** Transient reflectivity ($\Delta R/R$) spectra for $I_{ex}$ = 0.8 and 0.12 mJ/cm$^2$ at $t_d$ = 30 fs (closed blue circles for $I_{ex}$ =0.8 mJ/cm$^2$, open blue circles for $I_{ex}$ =0.12 mJ/cm²) and 300 fs (closed black circles for $I_{ex}$ = 0.8 mJ/cm²) after excitation by a 7 fs pulse. $\Delta R/R$ at $t_d$ = 300 fs after a 100-fs pulse excitation for $I_{ex}$ = 0.8 mJ/cm² is shown as the crosses. (b) Temporal change of the $\Delta R/R$ spectrum, plotted as a 2-dimensional (probe energy–delay time) map. Positive and negative $\Delta R/R$ are shown by the blue and red shadings, respectively.

Fig. 4 **Time profiles of transient reflectivity measured at 0.64 eV.**
(a) Time evolution of transient reflectivity ($\Delta R/R$) observed at 0.64 eV for $I_{ex}$ = 0.8 mJ/cm² (solid curve) and 0.12 mJ/cm² (dashed-dotted curve). The cross correlation between the pump and the probe pulses is also indicated by the orange shading. The dashed blue curve shows the positive component of $\Delta R/R$, reflecting the photoinduced charge localization, which was estimated assuming that the negative component grows exponentially (dashed red curve). (b) Oscillating component of the time profile. The time-resolved spectra of the oscillating component obtained by wavelet (WL) analysis (blue curve for 0–40 fs, red curve for 80–120 fs) are shown in the inset. Optical conductivity spectra at 10 K (CO) and 140 K (metal) are shown by the black curves in the inset.

Fig. 5 **Electric field and temperature dependences of transient reflectivity**
(a) Transient reflectivity ($\Delta R/R$) at 0.64 eV as a function of electric field $E_0$ ($\propto \sqrt{I_{ex}}$) for various temperatures. Inset: temperature dependence of $\Delta R/R$ at $E_0$ = 11.7 MV/cm. (b) Positive component of $\Delta R/R$ derived by subtracting the negative component (see text) as a function of $E_0$, together with $1/t'$ in units of $t_0$ ($E_0$ <13 MV/cm) calculated from equation 1 (solid red curve). **(c)** Zeroth-order Bessel function $J_0 \propto t'$ as a function of $E_0$.

**Fig. 6   Changes of transition temperature by reducing transfer integral .**

**(a)** The hole densities ($\rho_H$) at the A and A' molecules in Fig. 1(a) as a function of the normalized temperature $T/T_{CO}$, calculated using the Hartree–Fock approximation for an extended Hubbard model[36] for $t_0$, $0.95t_0$, $0.9t_0$, and $0.85t_0$ (circles, squares, triangles, and inverted triangles, respectively). **(b)** Change in $T_{CO}$ (= $\Delta T_{CO}/T_{CO}$) as a function of decreasing transfer integral $t$ ($-\Delta t/t$) shown by the black closed circles. The dashed line serves as a guide to the eye. The dashed red arrows indicate that a 10 % reduction in $t$ causes a 12 % increase in $T_{CO}$.

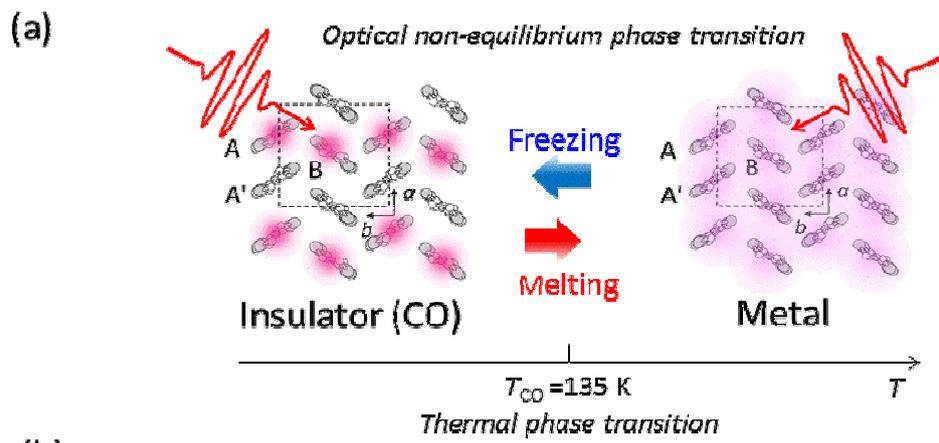

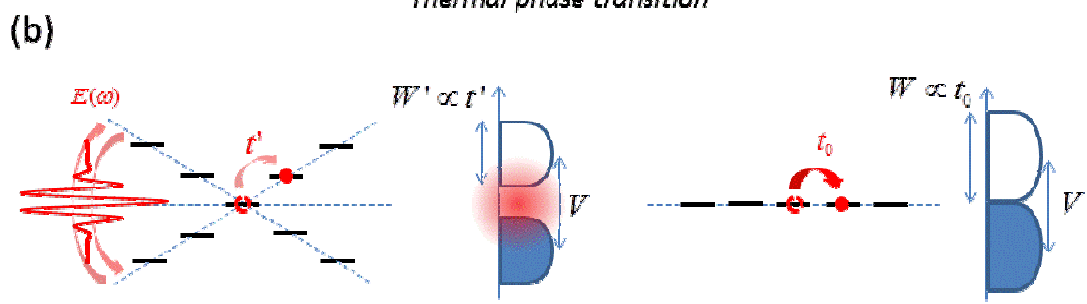

Ishikawa et al. Fig.1

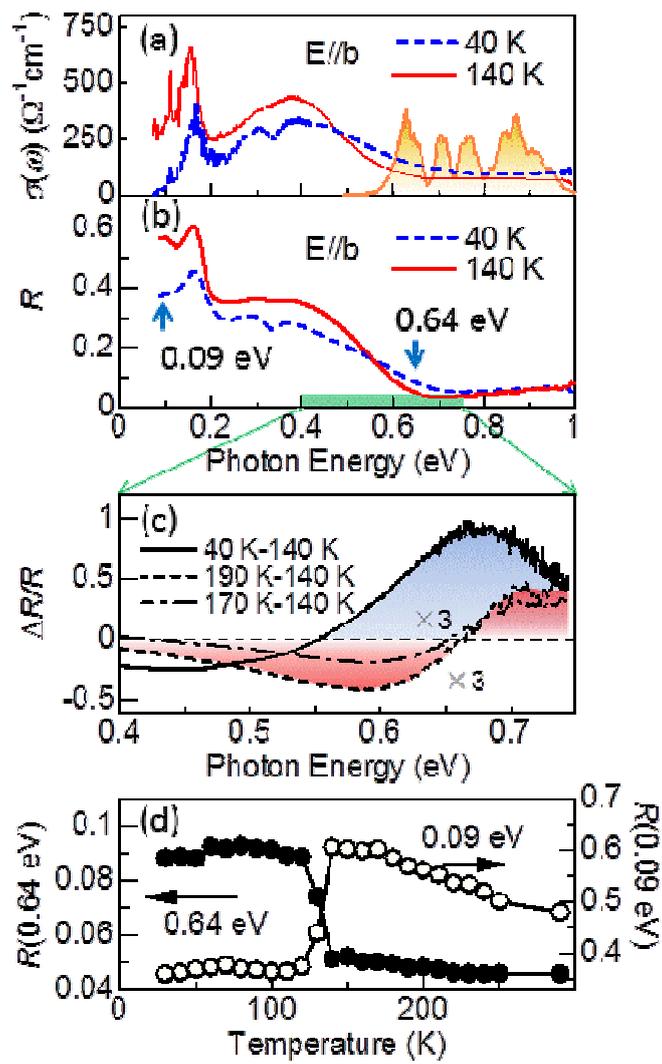

Ishikawa et al. Fig.2

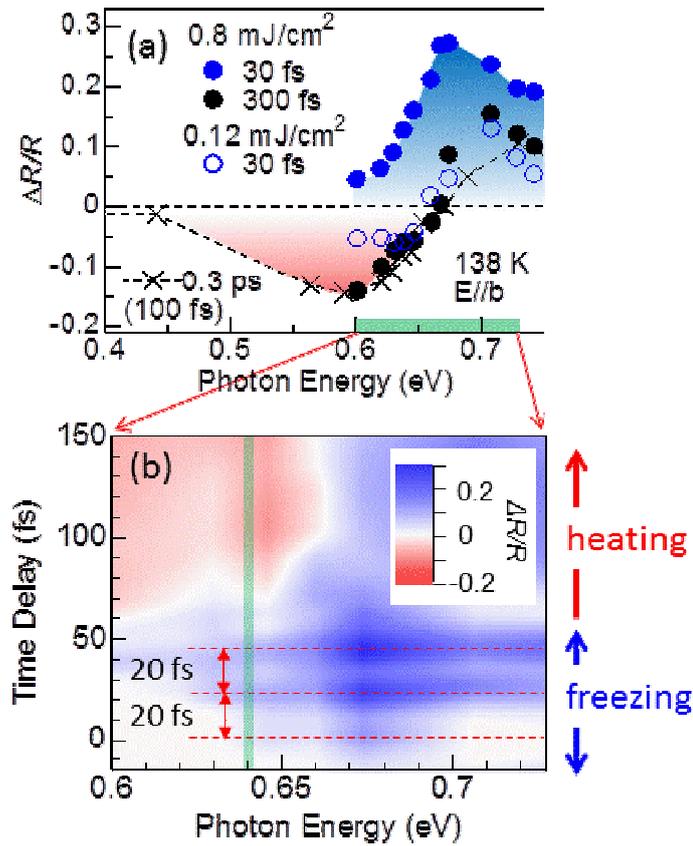

Ishikawa et al. Fig.3

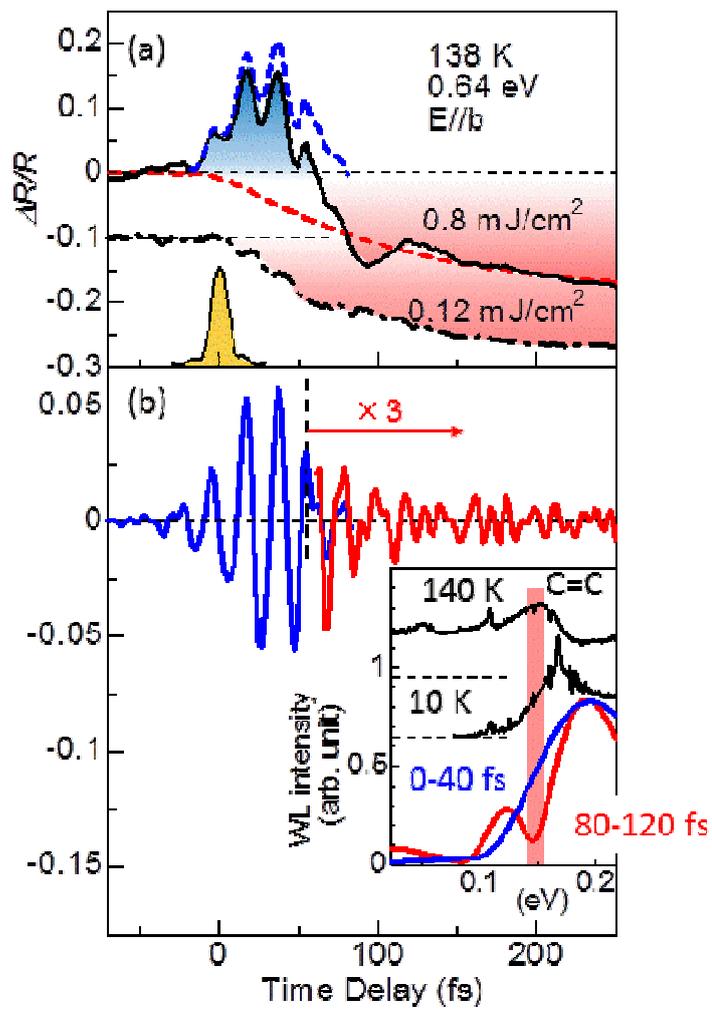

Ishikawa et al. Fig.4

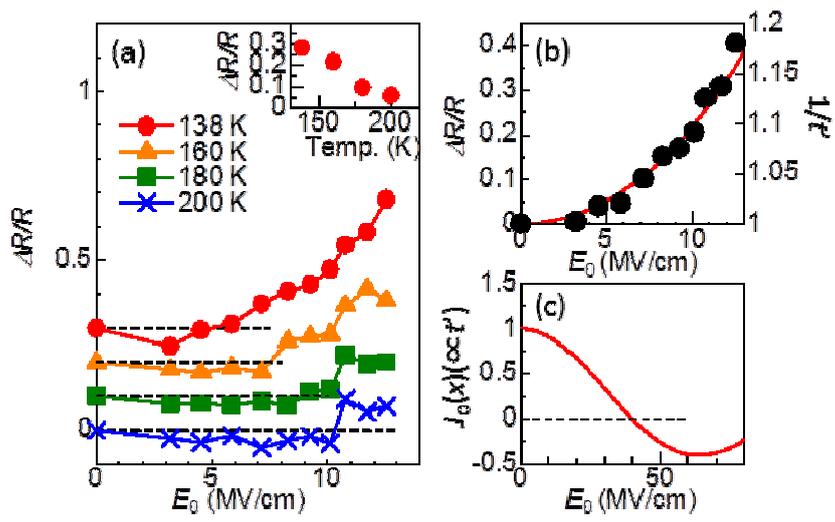

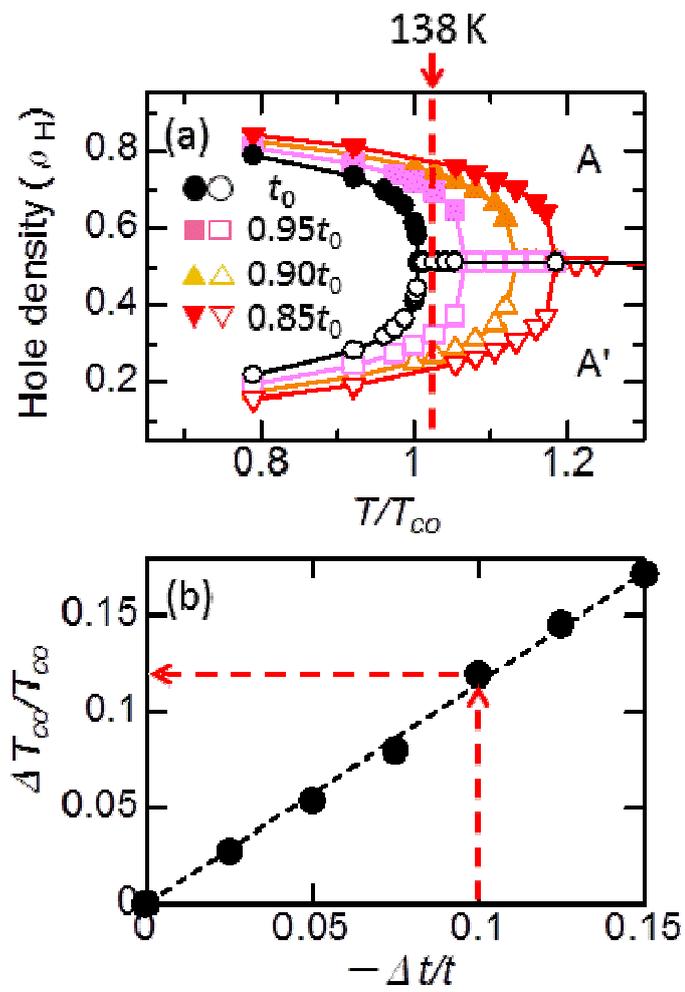